Realization of Carrier Tunneling from InAlAs Quantum Dots to AlAs

Masataka Koyama[1], Dai Suzuki[1], Xiangmeng Lu[1], Yoshiaki Nakata[2], and Shunichi Muto[1*]

[1]Department of Applied Physics, Graduate School of Engineering, Hokkaido University, Sapporo, Hokkaido, 080-8628, Japan

[2]Fujitsu Laboratories Ltd., 10-1 Morinosato-Wakamiya, Atsugi, 243-0197, Japan

*E-mail: GFD01102@nifty.com

Abstract

With the aim of improving solar cell efficiency, a structure for realizing electron tunneling from $In_{0.6}Al_{0.4}As$ quantum dots (QDs) through an $Al_{0.4}Ga_{0.6}As$ barrier to AlAs has been grown using molecular beam epitaxy. The photoluminescence decay time decreased from 1.1 ns to 390 ps as the barrier thickness decreased from 4 to 2 nm, which indicates that the photo-excited carriers tunneled from the QDs to the AlAs X energy level for a barrier thickness 2 nm in 0.6 ns, which is significantly longer than the tunneling time of GaAs and InAlAs quantum wells. We expect that this structure will assist in developing high-efficiency QD sensitized solar cells.

1. Introduction

Semiconductor quantum dots (QDs) capable of absorption over a wide wavelength range are ideal candidates for solar-cell photon absorbers and mid-gap absorbers for two photon absorption.[1] The fast recombination time as fast as sub-nanoseconds, however, is an obstacle for obtaining high efficiency photoelectric conversion. One method for improving the efficiency is to invoke carrier tunneling to cause the spatial separation of electrons and holes in the quantum dots.

In this paper, we describes a way to realize carrier tunneling from InAlAs QDs, which exhibit absorption in the visible region, through an AlGaAs barrier to the X level of AlAs. Interestingly, this structure simulates the structure of colloidal QD-sensitized solar cells (QDSSCs). Here, AlAs replaces indirect gap $TiO_2$, InAlAs QDs replace the core of colloidal QDs and AlGaAs replaces the shell of colloidal QDs.

After more than two decades of development, photosensitization of wide-band gap nano-crystalline semiconductors by adsorbed dyes has become a realistic option for solar cell applications, and dye-sensitized solar cells (DSSC), also known as Grätzel cells[2] currently present one of the most promising alternative to the conventional solar cells.[2-7] In a porous film consisting of nanometer-sized $TiO_2$ grains, the effective surface area for dye adsorption can be greatly enhanced and efficient light absorption is achieved from a dye molecule. For DSSCs, an overall energy conversion efficiency of 11.0% has been reported.[8,9] Because organic dye is instable in air, some researchers have proposed replacing the dye with colloidal nano-particles with core-shell.[10,11] However, the energy conversion efficiency is still low. We expect that the ideal carrier transport seen in our epitaxial structure will contribute to improve the efficiency of QDSSCs.

Self-assembled $In_xAl_{1-x}As$ QDs exhibiting absorption and emission in the visible range have been grown and studied by Fafard and coworkers.[12-16] The In concentration dependence of the surface coverage at the transition between 2D and 3D growth and the QD densities, and size distributions were observed by transmission electron microscopy.[13] Cathodeluminescence[14] and magnetoluminescence[15] were also completed and the recombination time of the carriers was estimated using resonant time resolved photoluminescence (TRPL).[16] However, there has been less research on InAlAs QDs than on InAs QDs, although InAlAs has become important for the single-dot devices applications because of its absorption and emission in the wavelength range for highly sensitive Si photodetectors.[17-21]

## 2. Experimental

The InAlAs/AlGaAs/AlAs tunneling structure was fabricated with a commercially available conventional molecular beam epitaxy (MBE) system with solid source K-cells. The sample was grown on a semi-insulating GaAs substrate adhered by In to a molybdenum block attached to the manipulator with tantalum planar heater. The growth chamber was cooled with liquid nitrogen, and the background pressure was kept below $10^{-9}$ Torr during growth.

The sample structure, illustrated in Fig. 1, contains two self-assembled $In_xAl_{1-x}As$ /$Al_{0.3}Ga_{0.7}As$ QD layers: one on the top surface for atomic force microscope (AFM) observation and the other sandwiched between two AlAs/AlGaAs layers for the photoluminescence (PL) measurements. A 400 nm undoped GaAs layer was grown epitaxially at 600 °C under a constant $As_4$ flux of $6 \times 10^{-6}$ Torr as a buffer layer, and the growth was stopped until the substrate temperature lowered to 490 °C under the same $As_4$ flux of $6 \times 10^{-6}$ Torr in 3 min. Then, GaAs (50 nm), AlAs (10 nm) and $Al_{0.3}Ga_{0.7}As$ (50 nm) layers followed by the $In_xAl_{1-x}As$ QDs were grown on the substrate. The $In_xAl_{1-x}As$ QD layer was grown by alternating the growth of sub-monolayer InAs and AlAs to obtain the desired In concentration, and the Stranski-Krastanow growth mode was employed: a layer is pseudomorphically grown at the initial growth stage on $Al_{0.3}Ga_{0.7}As$ to form the so-called wetting layer (WL), and the QDs are then formed by strain-relaxation into coherent islands. The transition from 2D to 3D growth was observed by a change in the reflection high energy electron diffraction (RHEED) pattern from a streaky to spotty pattern. The QDs were post-annealed for 1min at 490 °C under an $As_4$ flux of $6 \times 10^{-6}$ Torr. Further $Al_{0.4}Ga_{0.6}As$, AlAs, GaAs, and $Al_{0.3}Ga_{0.7}As$ layers were grown before the $In_xAl_{1-x}As$ QD layer for the AFM observations are grown.

The samples were observed with an AFM with a Si cantilever. The PL was excited by 15 mW continuous wave (CW) Ar ion laser with wavelength of 514 nm. PL spectra were taken when the sample was cooled to 18K. The TR-PL was obtained by excitation from SHG of 80 MHz pulsed titanium sapphire laser when the sample were cooled to 77 K.

## 3. Results and Discussions

Figure 2 shows the CW-PL spectrum with varied In contents. QD growths were stopped just after RHEED pattern became sufficiently spotty. A broad PL emission peak is observed due to size fluctuation of the $In_xAl_{1-x}As$ QDs. The In concentration dependence of the peak energies is shown in Fig. 3. As expected from the band gap of InAlAs, the peak energy is red-shifted as the In concentration is increased. However, it should be noted that the red-shift is the result of change in the stress in the QD and the coverage in addition to the unstrained band gap of QD material.

Figure 4 shows a typical AFM image for an $In_{0.6}Al_{0.4}As$ coverage of 4.1 monolayer (ML). The average height of the QDs was 0.6 nm with an average diameter and density of 41.5 nm and $4.5 \times 10^9$ $cm^2$, respectively. The average QD diameters and heights at different coverages are also shown in Fig. 4.

Figure 5 shows the PL peak wavelength and the full width at half maximum (FWHM) of $In_{0.6}Al_{0.4}As$ at different coverages. As was seen in Fig. 4, a higher dot height resulted in higher coverage and the PL peak shifts to a longer wavelength in accordance with the quantum size effect.

Figure 6 shows the structure we grew for the tunneling experiment. The carriers are expected to tunnel from the $In_{0.6}Al_{0.4}As$ QD layers through $Al_{0.4}Ga_{0.6}As$ barrier to AlAs X-band. We fabricated samples with $Al_{0.4}Ga_{0.6}As$ tunneling barrier with thickness $L_B$ = 2 and 4 nm and $In_{0.6}Al_{0.4}As$ QD layers of coverage 5.2ML. For reference, we grew samples without QDs with $L_B$ of 8 nm and 2 nm and $In_{0.6}Al_{0.4}As$ wetting layer (WL) of 3.8 ML.

Figure 7 shows 2D mapping of the TR-PL intensities for $L_B$ of 4 and 2 nm. As shown in Fig. 8, The decay time of $L_B$ = 2 nm is shorter than $L_B$ = 4 nm. Figure 9 is 2D mapping of the TR-PL intensities of WLs with $L_B$ = 8 nm and $L_B$ = 2 nm. As shown in Fig. 10, the decay times are much faster than QD samples .

A red shift and broadening is seen in 2 nm barrier compared to the 4 nm one in Fig. 7. A similar red shift is seen in Fig. 9 of the WL samples with decreasing $L_B$. This contrasts with a previous report on the GaAs quantum well (QW) sandwiched by AlAs through AlGaAs barriers.[22] They observed a slight blue shift as the $Al_{0.51}Ga_{0.49}As$ barrier was thinned from 3.4 to 1.1 nm, which was attributed to the increased confinement by the Γ level of AlAs with decreasing the AlGaAs barrier. The reason for the present spectral change with decreasing the $L_B$ is not clarified yet. However, the slight change of the lattice constant between AlGaAs and AlAs could have influenced the shape and its distribution of QDs.

The relaxation times estimated from the TR-PL spectra are 1.1 ns and 390 ps for barrier thicknesses of 4 and 2 nm, respectively. The carrier recombination time of InAlAs QDs were reported to be 0.5 ns or longer[17-21] for similar emission energies. Therefore, a relaxation time of 390 ps for the 2 nm barrier can be regarded to have effects of electron tunneling. The PL decay is expected to consist of two processes: recombination of electron-hole pairs in the QDs with a decay time of $\tau_{rec}$ and electron tunneling through the AlGaAs barrier to the AlAs X level with decay time of $\tau_{tunnel}$

The decay time $\tau_{decay}$ can then be expressed as

$$1/\tau_{decay}=1/\tau_{tunnel}+1/\tau_{rec}.$$

The observed 1.1 ns decay of the 4 nm barrier is considered to have negligible contribution from tunneling and can be attributed solely to recombination. Using the value of 1.1 ns for $\tau_{rec}$ and the observed decay time of 0.39 ns, we obtained $\tau_{tunnel}$ = 0.60 ns.

Figure 11 compares the observed decay times with those reported for GaAs QW (2.8 nm)/Al$_{0.51}$Ga$_{0.49}$As/AlAs (7.1 nm) in which the carrier transfer was identified as electron tunneling from the GaAs Γ level to the AlAs X level.[22] QW results based on the InAlAs WL in our structure are also shown. The QD result has a much slower tunneling time than the GaAs and InAlAs QWs, and while the tunneling mechanism for these structures is yet to be identified, it could be attributed to phonon-assisted tunneling together with Γ-X transfer by Γ-X mixing at the interface and direct tunneling together with Γ-X intervalley scattering such as by intervalley phonons. Therefore, there can be two reasons for the slow tunneling time of the QDs. Since the PL peak energy (0.75 μm) is smaller than that of the GaAs QWs (0.7 μm), the electron energy of the QDs measured from the minima of the AlAs X level is expected to be smaller than that of the QWs assuming the constant ratio of energy difference between electron and hole. This could result in suppressed emission of high energy phonons, such as longitudinal optical (LO) phonons and intervally phonons, important for phonon-assisted tunneling. There is a dimensional mismatch of wave functions across the tunneling barrier between the 0D of QD and the 2D or quasi-3D of AlAs layer. This could also result in slower tunneling due to reduced overlapping of wave functions before and after tunneling.

Although we gave straightforward interpretation above, we also have to consider possibilities of unexpected mechanisms other than tunneling. First of all, as we decrease $L_B$, we are decreasing the distance to the AlAs layer. Therefore, although not likely, if AlAs is contaminated somehow, it can cause some unwanted decay mechanism. One of them is the introduction of nonradiative center within QDs such as vacancies and deep impurities. However, these effects are expected to result in the same decay rate both for the QDs and WL. Therefore, the observed difference in decay time between QD and WL shown in Fig. 11 excludes this possibility. The other effect of AlAs is the presence of nonradiative center within AlAs or at AlAs/AlGaAs interface. This is also a part of tunneling and we do not need to exclude this. However, for solar cell applications, we have to erase this in the future if it should exist.

We don't know the reason for the difference of decay times between WL and GaAs QW shown in Fig. 11. They are both 2 dimensional and we cannot attribute the difference to dimension. As already discussed, the slight change of energy difference, before and after tunneling is critical for emission of LO and intervalley phonons and can drastically influences the tunneling time. Therefore, this could be a part of the difference of decay times. Hence, a precise measurement of energy difference is required for further confirmation of tunneling.

## 4. Conclusions

A structure for promoting carrier tunneling from the $In_{0.6}Al_{0.4}As$ QDs through an $Al_{0.4}Ga_{0.6}As$ barrier to the AlAs X level has been grown by MBE. The PL decay time was shown to decrease from 1.1 ns to 390 ps as the barrier thickness decreased from 4 to 2 nm, indicating that photoexcited electrons were extracted by tunneling from the QDs to the AlAs X level for the 2 nm barrier structure. The results suggest that even 2 nm AlGaAs barriers is not thin enough to tunnel compared with recombination and that further understanding is necessary to realize the effective separation of carriers.

Figure captions

Fig. 1. (Color online) MBE grown structure for AFM and PL.

Fig. 2. (Color online) Photoluminescence of $In_xAl_{1-x}As/Al_{0.3}Ga_{0.7}As$ QD at 18K. The sharp peaks are emission from GaAs QWs (30nm).

Fig. 3. (Color online) The In concentration x-dependence of PL peak wavelength of $In_xAl_{1-x}As/Al_{0.3}Ga_{0.7}As$ QD.

Fig. 4. (Color online) Typical AFM image of QDs (left) and the coverage dependence of average diameter and height of $In_{0.6}Al_{0.4}As$ QDs from AFM images.

Fig. 5. (Color online) Coverage dependence of photoluminescence wavelength and FWHM of $In_{0.6}Al_{0.4}As$ QDs (coverage: 4.1, 5.2, and 6.2 ML).

Fig. 6. (Color online) Tunneling structure for TR-PL. Electrons can tunnel from $In_{0.6}Al_{0.4}As$ QDs through $Al_{0.4}Ga_{0.6}As$ ($L_B$) to AlAs (10 nm) layer beneath.

Fig. 7. (Color online) Streak Camera mapping for QDs with $L_B$ = 4 and 2 nm at 77 K. Pumping Source: Titanium Sapphire Laser with average power 2 mW. Excitation Source: SHG 448 nm.

Fig. 8. (Color online) Intensity decay curves for barrier thickness $L_B$ = 4 and 2 nm by time resolved photoluminescence.

Fig. 9. (Color online) Streak camera mapping of TR-PL from WLs with $L_B$ of 8 and 2 nm. Excitation condition is the same as Fig. 7.

Fig. 10. (Color online) Intensity decay curve for WLs with $L_B$ = 8 and 2 nm.

Fig. 11. (Color online) Decay times for QDs (▲) and WLs (■). Also shown are the QW results (♦) from ref. 22 where tunneling from GaAs QW (2.8 nm) through $Al_{0.51}Ga_{0.49}As$ barrier ($L_B$) to AlAs(7.1 nm) was reported. Titled lines are eye guides drawn pararell to the one by ref. 22.

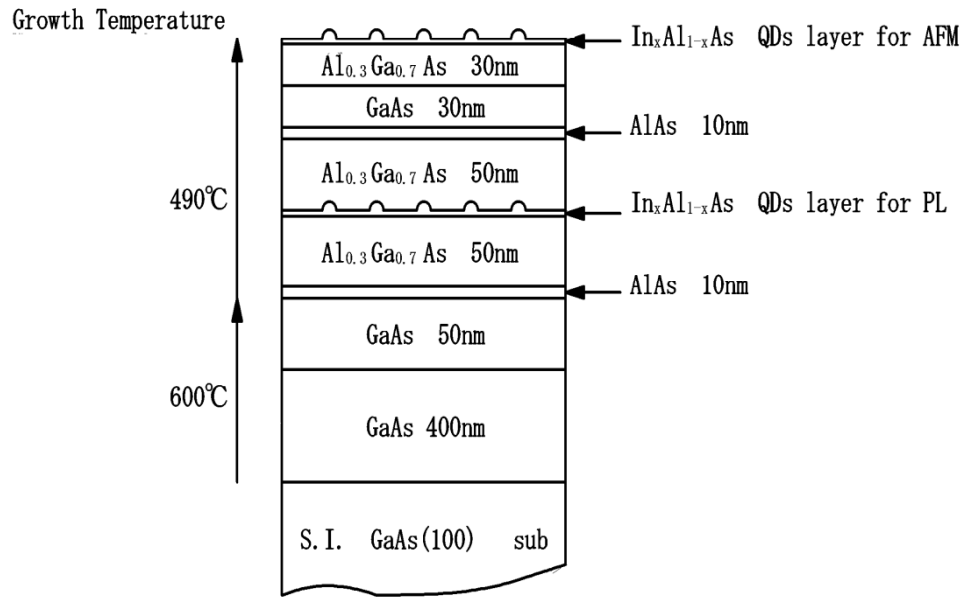

Fig. 1. (Color online) MBE grown structure for AFM and PL.

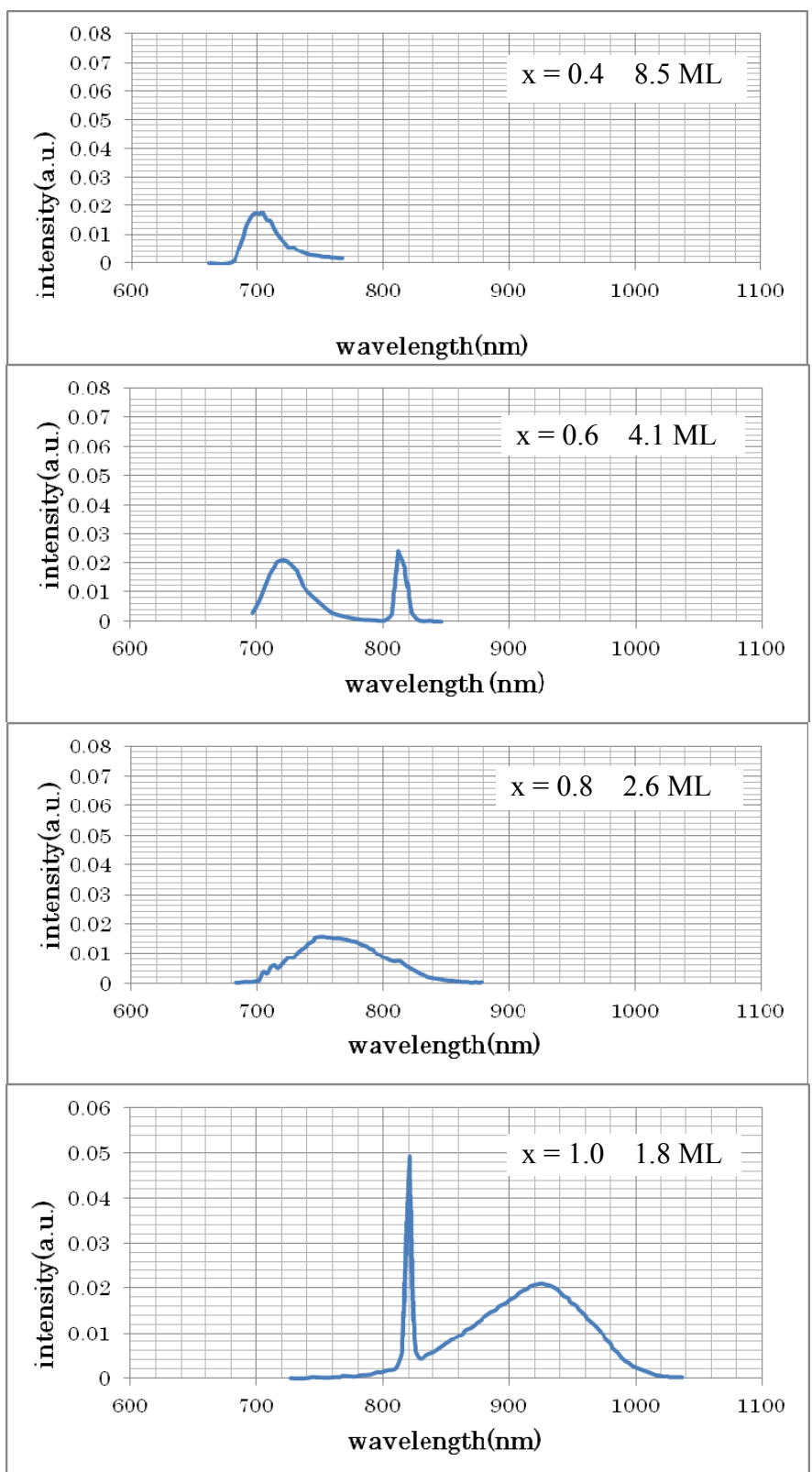

Fig. 2. (Color online) Photoluminescence of $In_xAl_{1-x}As/Al_{0.3}Ga_{0.7}As$ QD at 18K. The sharp peaks are emission from GaAs QWs (30nm).

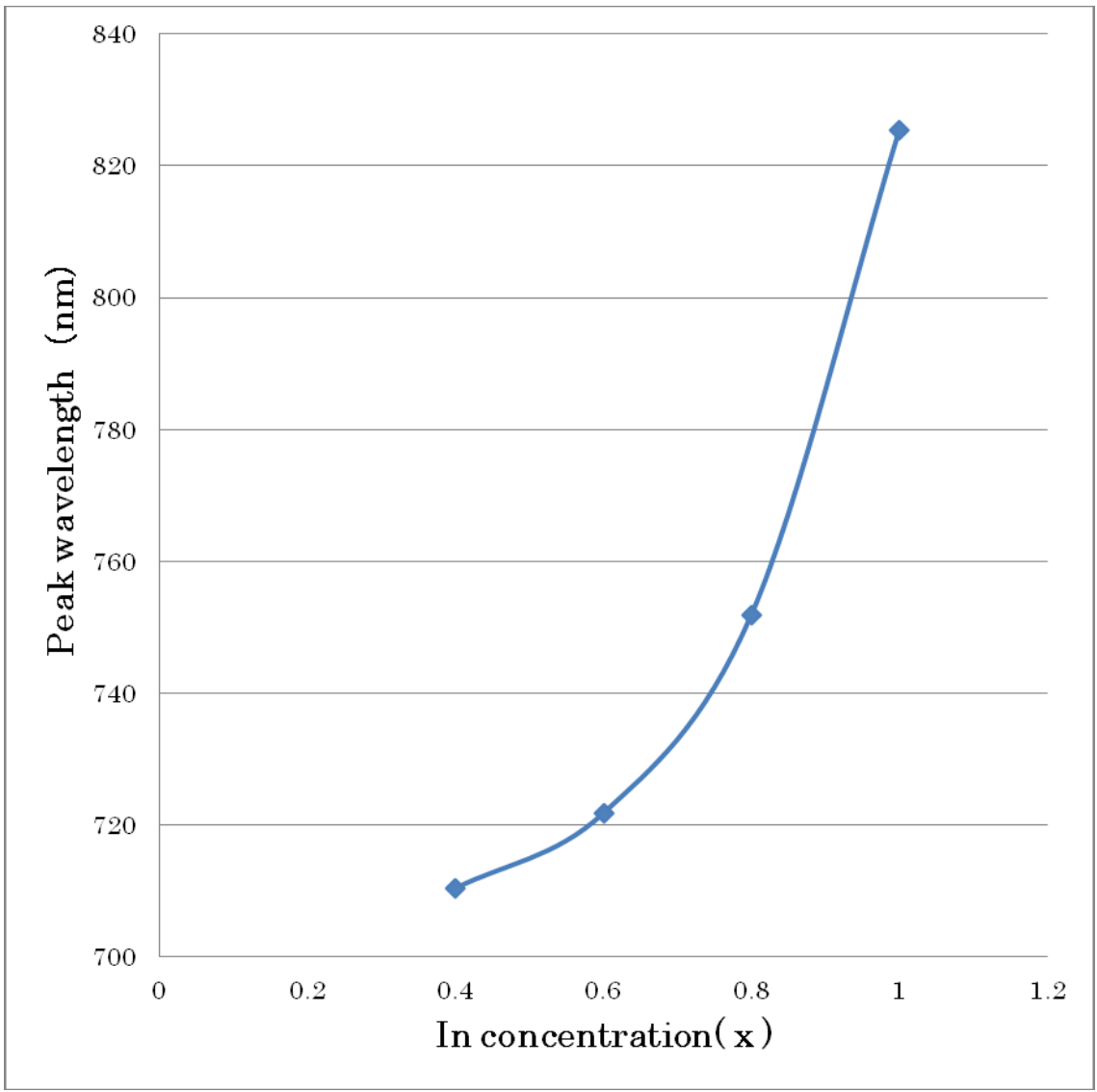

Fig. 3. (Color online) The In concentration x-dependence of PL peak wavelength of $In_xAl_{1-x}As/Al_{0.3}Ga_{0.7}As$ QD.

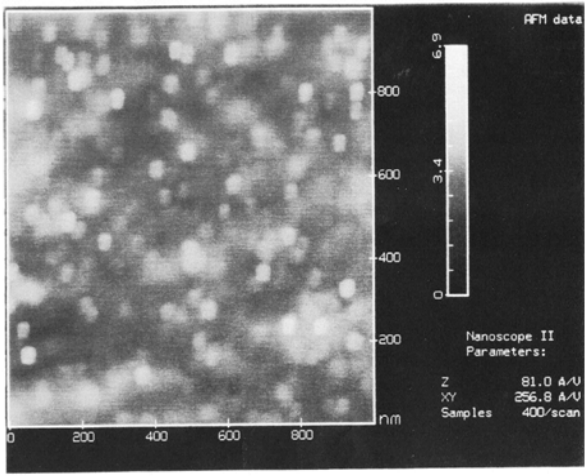 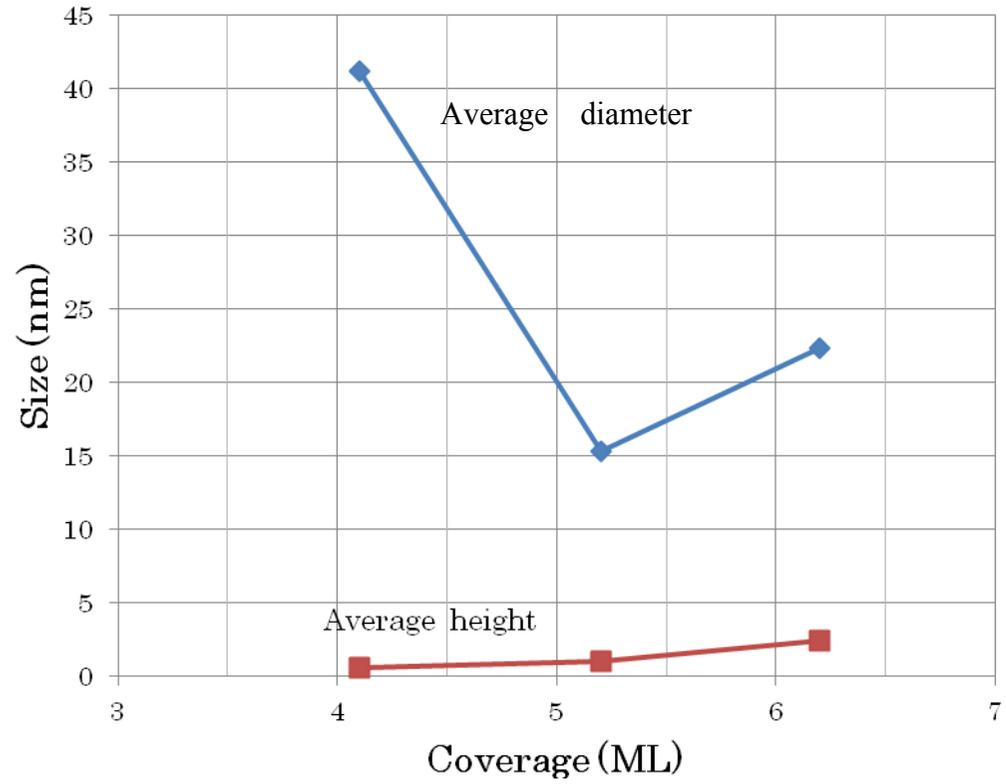

Fig. 4. (Color online) Typical AFM image of QDs (left) and the coverage dependence of average diameter and height of $In_{0.6}Al_{0.4}As$ QDs from AFM images.

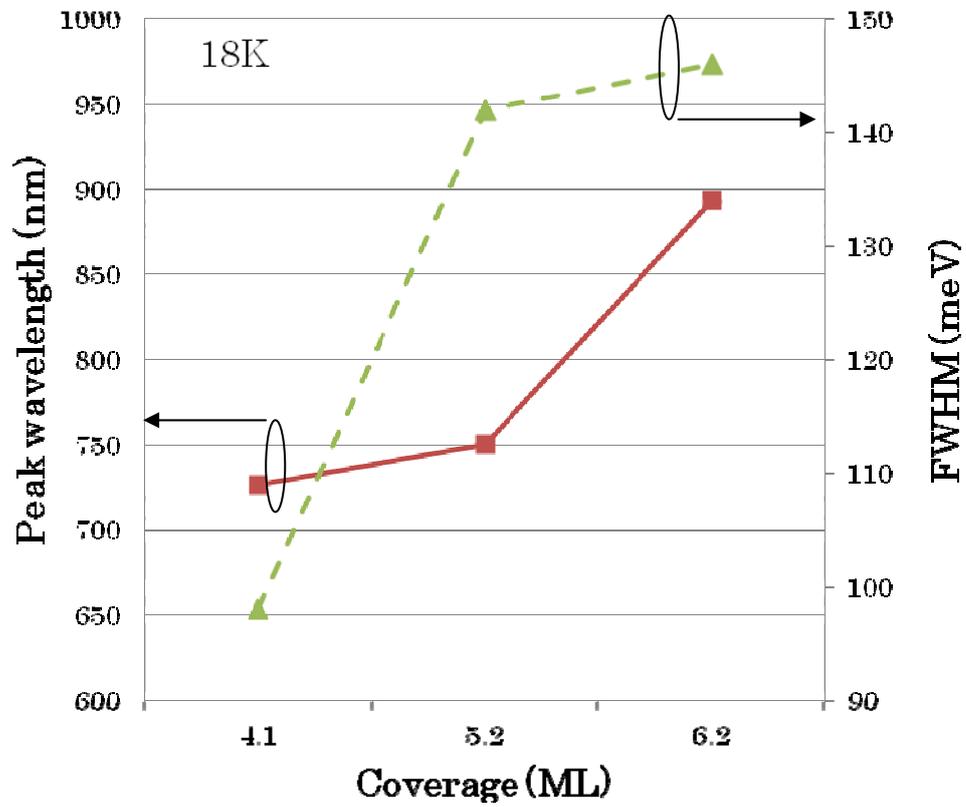

Fig. 5. (Color online) Coverage dependence of photoluminescence wavelength and FWHM of $In_{0.6}Al_{0.4}As$ QDs (coverage: 4.1, 5.2, and 6.2 ML).

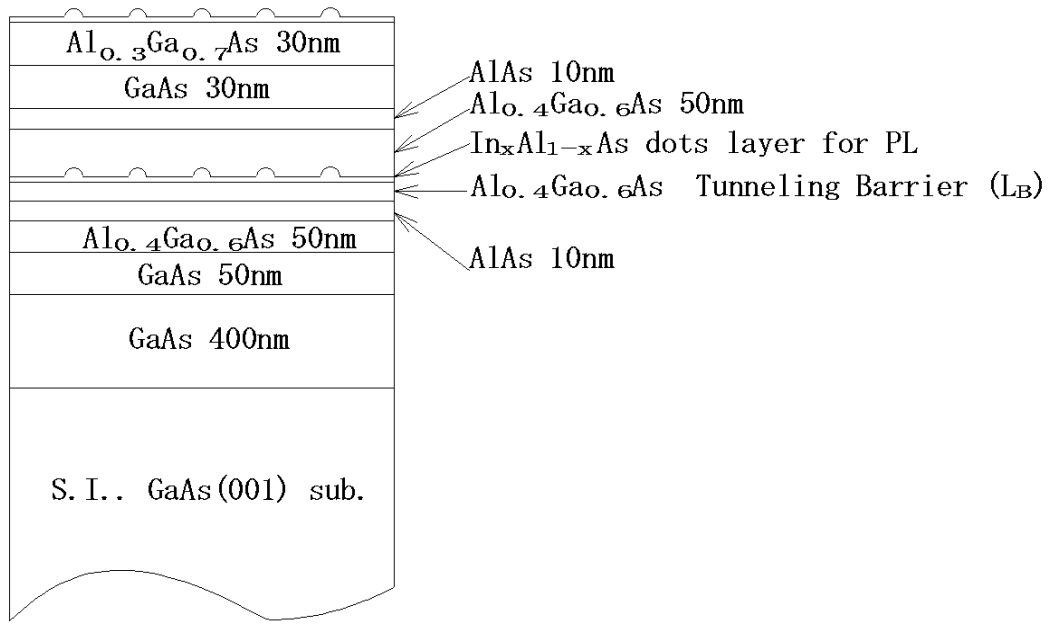

Fig. 6. (Color online) Tunneling structure for TR-PL. Electrons can tunnel from $In_{0.6}Al_{0.4}As$ QDs through $Al_{0.4}Ga_{0.6}As$ ($L_B$) to AlAs (10 nm) layer beneath.

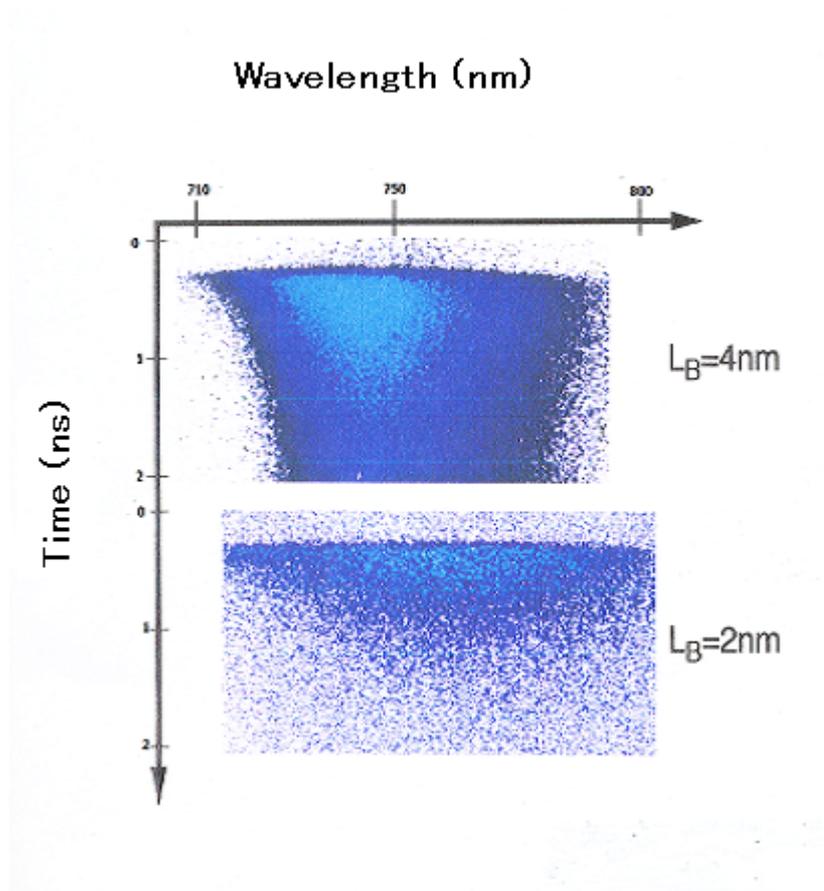

Fig. 7 (Color online) Streak Camera mapping for QDs with $L_B$ = 4 and 2 nm at 77 K. Pumping Source: Titanium Sapphire Laser with average power 2 mW. Excitation Source: SHG 448 nm.

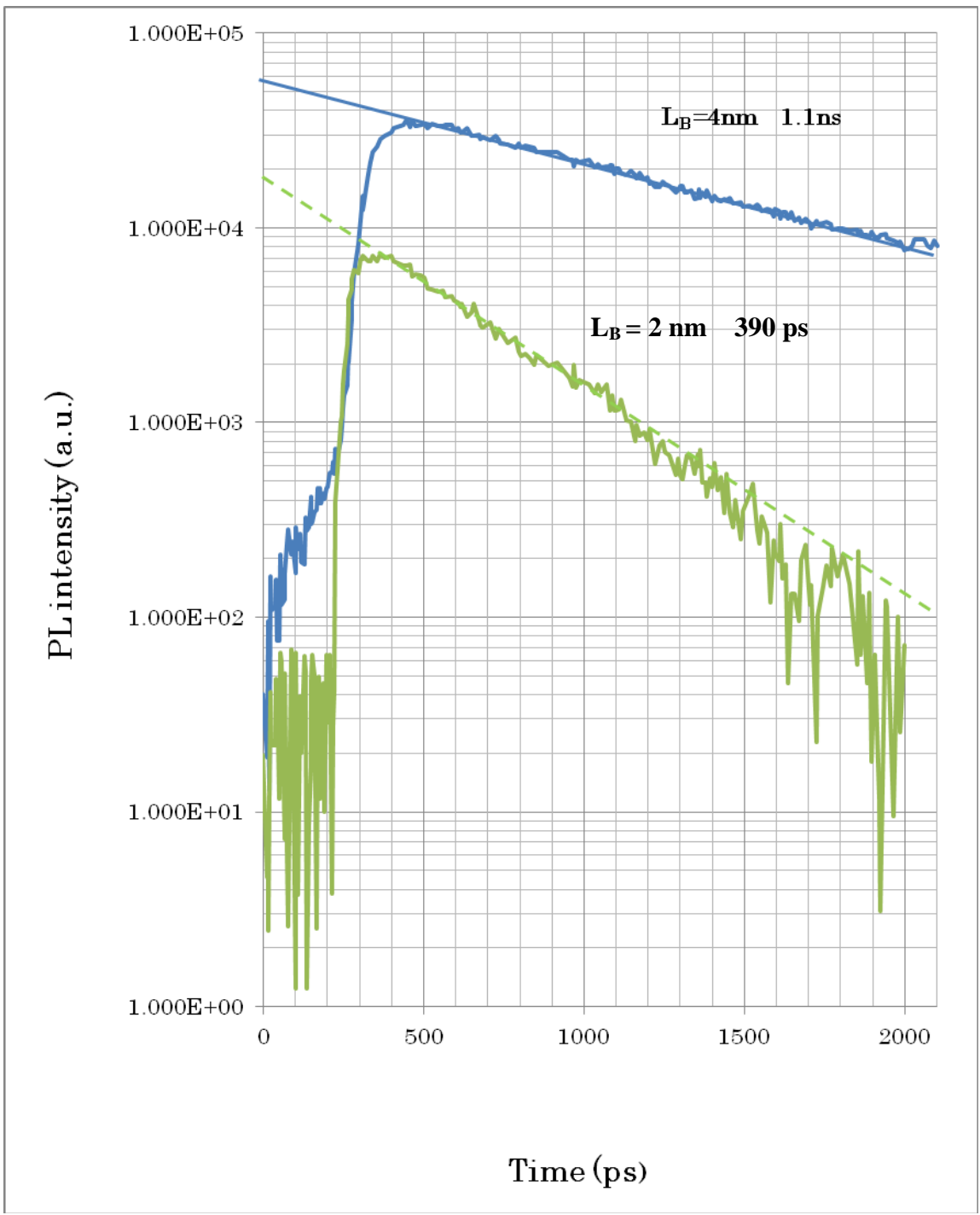

Fig. 8. (Color online) Intensity decay curves for barrier thickness $L_B$ = 4 and 2 nm by time resolved photoluminescence.

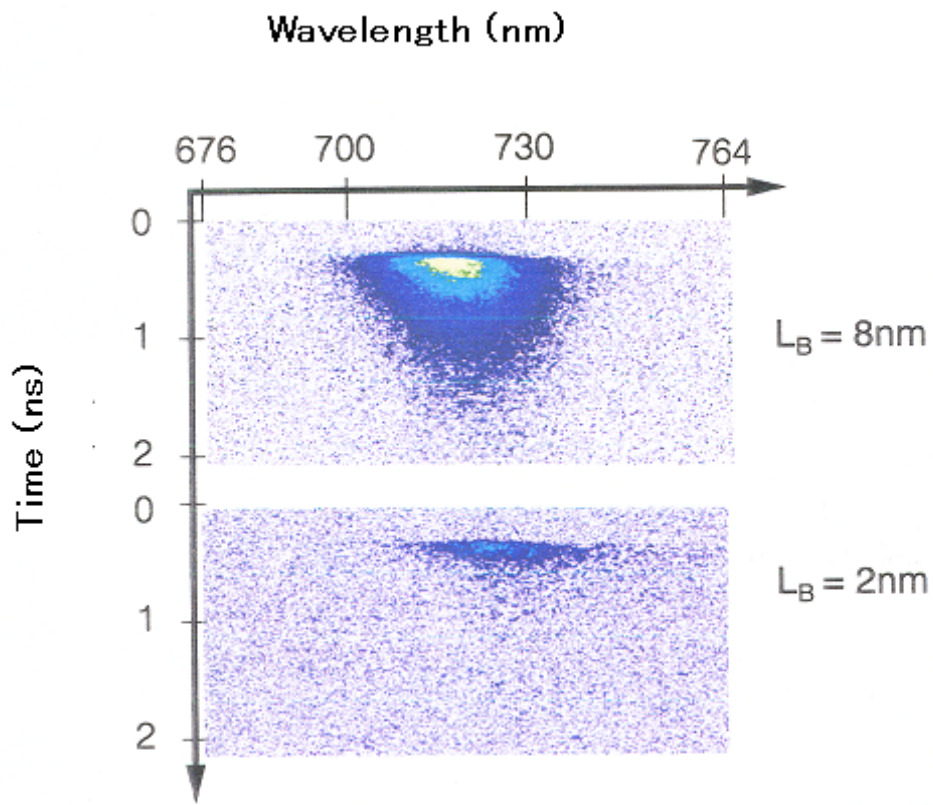

Fig. 9. (Color online) Streak camera mapping of TR-PL from WLs with $L_B$ of 8 and 2 nm. Excitation condition is the same as Fig. 7.

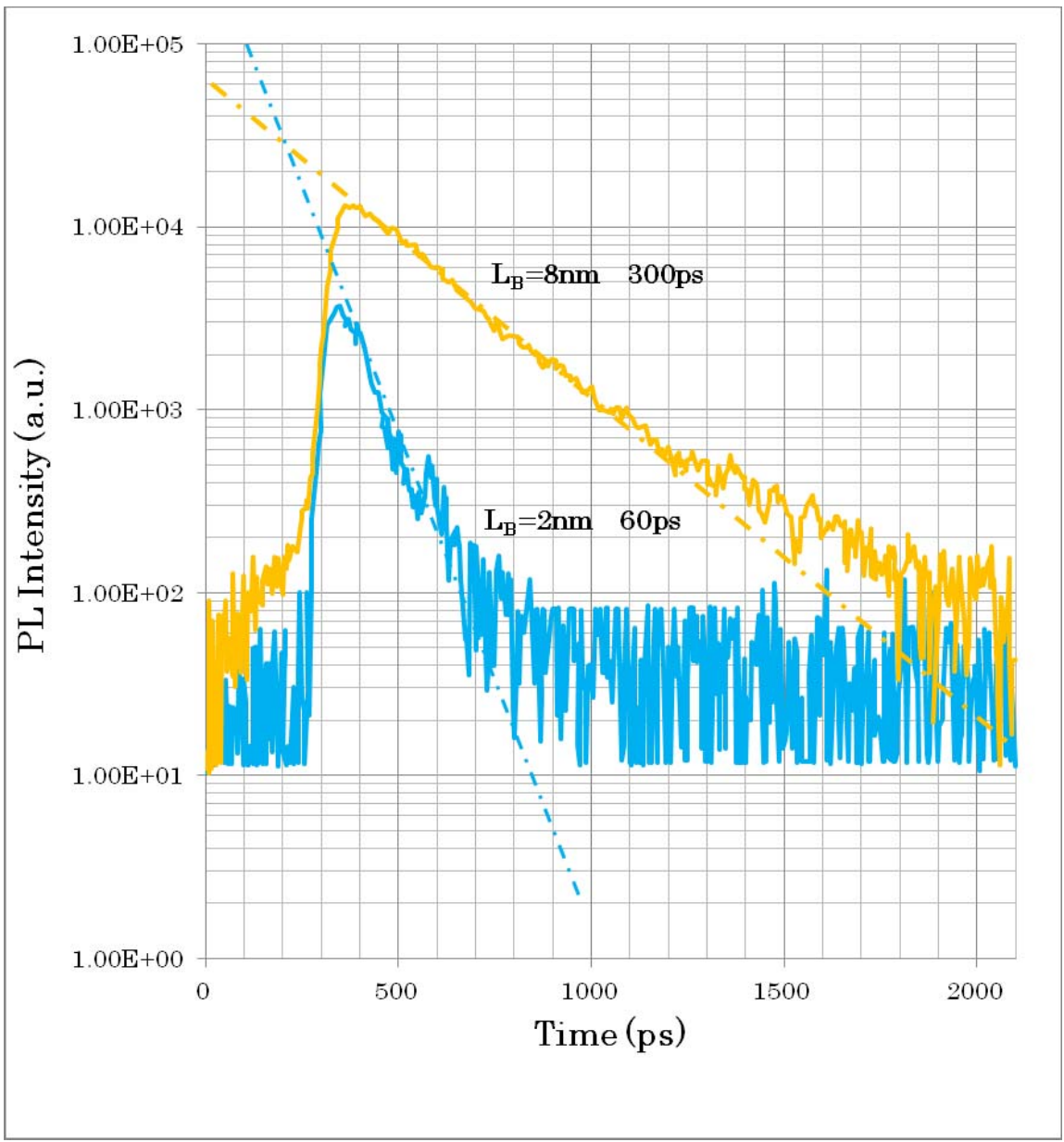

Fig. 10. (Color online) Intensity decay curve for WLs with $L_B$ = 8 and 2 nm.

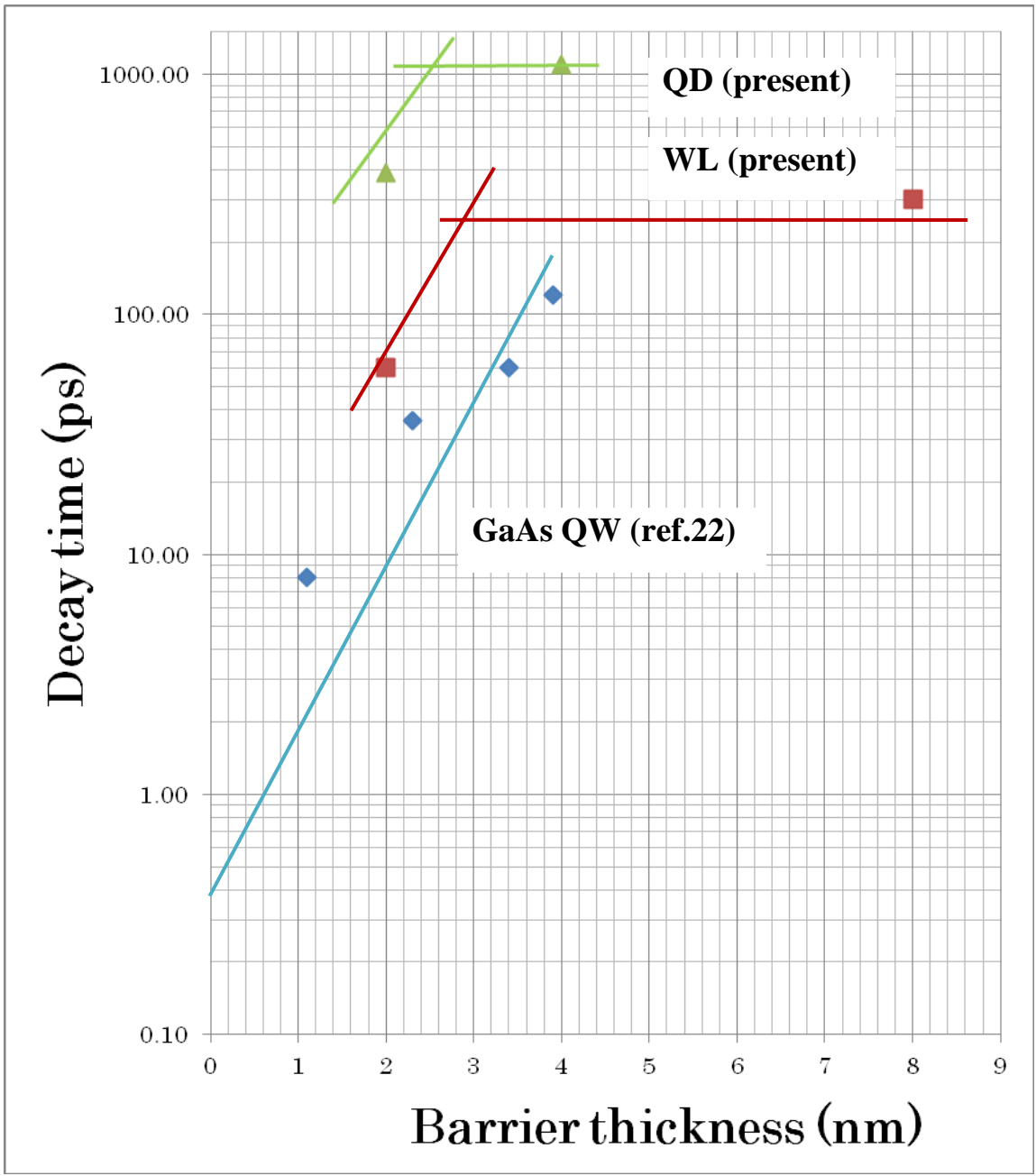

Fig. 11. (Color online) Decay times for QDs (▲) and WLs (■). Also shown are the QW results (♦) from ref. 22 where tunneling from GaAs QW (2.8 nm) through $Al_{0.51}Ga_{0.49}As$ barrier ($L_B$) to AlAs(7.1 nm) was reported. Titled lines are eye guides drawn pararell to the one by ref. 22.

26